\documentclass[useAMS,usenatbib]{mn2e}

\usepackage{lscape}
\usepackage{graphicx}
\usepackage{natbib}
\usepackage{graphics}

\usepackage{amsmath}

\title[The evolution of galaxy local densities]{The evolution of the density of galaxy clusters and groups:\\
denser environments at higher redshifts.
}

\author[Bianca M. Poggianti et al.]{\parbox[t]{\textwidth}{Bianca M. Poggianti$^1$\thanks{E-mail: bianca.poggianti@oapd.inaf.it}, Gabriella De Lucia$^2$, Jesus Varela$^1$, Alfonso Aragon-Salamanca$^3$, Rose Finn$^4$, Vandana Desai$^5$, Anja von der Linden$^6$, Simon D.M. White$^7$.}\\
\\
$^1$INAF-Astronomical Observatory of Padova, Italy\\
$^2$INAF-Astronomical Observatory of Trieste, Italy\\ $^3$School of
Physics and Astronomy, University of Nottingham, United Kingdom\\
$^4$Department of Physics, Siena College, Loudonville, USA\\
$^5$Spitzer Science Center, California Institute of Technology, USA\\
$^6$Kavli Institute for Particle Astrophysics and Cosmology, Stanford University, USA\\ $^7$Max-Planck-Institut fur Astrophysik, Garching,
Germany}

\begin{document}

\date{Accepted .... Received .....; in original form .....}

\pagerange{\pageref{firstpage}--\pageref{lastpage}} \pubyear{2009}

\maketitle
\label{firstpage}

\begin{abstract}

We show that, observationally, the projected local density
distribution in high-z clusters is shifted towards higher values
compared to clusters at lower redshift. To search for the origin of
this evolution, we analyze a sample of haloes selected from the
Millennium Simulation and populated using semi-analytic models,
investigating the relation between observed projected density and
physical 3D density, using densities computed from the 10 and 3
closest neighbours.  Both observationally and in the simulations, we
study the relation between number of cluster members and cluster mass,
and number of members per unit of cluster mass.  We find that the
observed evolution of projected densities reflects a shift to higher
values of the physical 3D density distribution.  In turn, this must be
related with the globally higher number of galaxies per unit of
cluster volume $N/V$ in the past. We show that the evolution of $N/V$
is due to a combination of two effects: a) distant clusters were
denser in dark matter (DM) simply because the DM density within
$R_{200}$ ($\sim$ the cluster virial radius) is defined to be a fixed
multiple of the critical density of the Universe, and b) the number of
galaxies per unit of cluster DM mass is remarkably constant both with
redshift and cluster mass if counting galaxies brighter than a
passively evolving magnitude limit.  Our results highlight that
distant clusters were much denser environments than today's clusters,
both in galaxy number and mass, and that the density conditions felt
by galaxies in virialized systems do not depend on the system mass.

\end{abstract}





\section{Introduction}

Ever since the work by Dressler et al. (1980) presenting the first
morphology-density relation in local galaxy clusters, the projected
number density of galaxies (the number of galaxies per unit area or
volume) has been a primary tool for investigating the relation between
galaxy properties and their environment.  Thirty years on, this tool
has been applied to galaxies in clusters, groups and the general
field, from the local to the high-z Universe, to study the systematic
variations of galaxy morphologies, colors, current star formation
activity, stellar masses and other galaxy properties with environment
(Postman \& Geller 1984, Dressler et al. 1997, Hashimoto et al. 1998,
Lewis et al. 2002, Gomez et al. 2003, Blanton et al. 2003, Kauffmann
et al. 2004, Balogh et al. 2004, Hogg et al. 2004, Postman et
al. 2005, Baldry et al. 2006, Cucciati et al. 2006, Cooper et
al. 2007, Elbaz et al. 2007, Cooper et al. 2008, Tasca et al. 2009,
Bolzonella et al. 2009, to name a few).

More and more sophisticated methods to characterize the ``local''
galaxy density have been scrutinized, and by now a variety of density
estimators have been employed in the literature, including the stellar
mass density (Wolf et al. 2009), the dark matter density field (Gray
et al. 2004), galaxy counts within a fixed metric projected radius and
a fixed recession velocity difference (Blanton et al. 2003, Hogg et al. 2004,
Kauffmann et al. 2004) and the projected $n$th-nearest neighbour
distance, with the optimal $n$ varying with environment and survey
characteristics, sometimes expressed as an overdensity with respect to
the median density in the survey (eg. see Cooper et al. 2005 and Kovac
et al. 2009 for thorough discussions of different methods). Widely
different methods explore different ``environmental'' conditions and
can probe very different physical scales, in some cases over several
Mpc, which are far from being a measure of the ``local''
environment as in the traditional, close neighbour studies in
clusters.

Nevertheless, all studies employing the galaxy density share the goal
of uncovering the environmental dependence of galaxy properties, and
have successfully proved it to be strong. In contrast, little
attention has so far been devoted to {\it how and why density
conditions change with redshift}. Observationally, density
distributions in clusters at different redshifts have never been
compared.  In the field, an overdensity is usually measured, factoring
out the evolution of the mean density of the Universe, and the
evolution of the distribution of environmental conditions beyond that
has never been investigated.  Theoretical efforts have so far focused
on uncovering the physical origin of the observed relations
between galaxy properties and density, or have been compared with
observations to test the model consistency, or to study the relation
between density and halo mass (e.g. Kauffmann et al. 2004, Berlind et
al. 2005, Baldry et al. 2006, Elbaz et al. 2008, Gonzalez \& Padilla
2009).  

A knowledge of the density distribution of galaxies as a function of
redshift, 
and as a function of system mass in virialized systems, would be very useful to
assess the changes of local environment experienced by galaxies, but
it is currently lacking. In this paper, we analyze the evolution of
galaxy densities in clusters, starting from the observed density
distribution in distant and nearby clusters. We employ the most
traditional of all density estimators, the 10 (or 3) closest neighbours
density, as in the original Dressler 1980 work and in most subsequent
cluster studies. We compare observations with Millennium Simulation
results in order to explain the observed change of the cluster density
distribution with redshift, and to relate the evolution in galaxy
number density with that of the matter density.

All cluster velocity dispersions $\sigma$ are given
in the rest frame. We assume a $\Lambda$CDM cosmology with
($H_0$, ${\Omega}_0$, ${\Omega}_{\lambda}$) = (73 $\rm km \, s^{-1} \, 
Mpc^{-1}$, 0.25, 0.75).

\section{Observations}

In this paper we exploit two samples of galaxy clusters: the ESO
Distant Cluster Survey (hereafter, EDisCS) at high-z, and a low-z
sample from the SDSS. We use the projected local galaxy densities
and the total number of members in these clusters as
key observables for our study.

The ESO Distant Cluster Survey (hereafter, EDisCS) is a photometric
and spectroscopic survey of galaxies in 20 fields containing galaxy
clusters at $z=0.4-1$. An overview of the goals and strategy of this
survey is given in White et al. (2005) who present the sample
selection and the optical ground--based photometry.  For all 20 fields
EDisCS has obtained deep optical photometry with FORS2/VLT, near-IR
photometry with SOFI/NTT, multislit spectroscopy with FORS2/VLT, and
MPG/ESO 2.2/WFI wide field imaging in $VRI$.  ACS/HST mosaic imaging
in $F814W$ of 10 of the highest redshift clusters has also been
acquired (Desai et al. 2007), and a number of multiwavelength follow-up
studies have been conducted (see Poggianti et al. 2009 for an overview).

Spectra of $>100$ galaxies per cluster field were obtained for 18 out
of the 20 cluster fields. The spectroscopic selection, observations,
and catalogs are presented in Halliday et al. (2004) and
Milvang-Jensen et al. (2008), together with the cluster velocity
dispersions. Spectroscopic completeness functions and [OII] line
properties as a function of cluster mass are discussed in Poggianti et
al. (2006). In this paper we use the projected local galaxy densities
computed by Poggianti et al. (2008), who present the star
formation-density and morphology-density relations. 
Our sample consists of 18 clusters and groups at
$z=0.4-0.8$ covering a wide range of velocity dispersions ($\sim 200 -
1100 \, \rm km \, s^{-1}$), as listed in Table~1.

At low-z, we use a compilation of 23 clusters and groups at
$0.04<z<0.1$ drawn from the SDSS, spanning a similar range of velocity
dispersions to EDisCS. The sample selection, completeness and data
used are fully described in Poggianti et al. (2006).  Projected local
densities were computed by Poggianti et al. (2008), in a similar way
as it was done for EDisCS (see \S4).

\begin{table}
\begin{minipage}{250mm}
\caption{List of clusters.\label{tbl1}}
\begin{tabular}{llcl}
\hline\hline
&&& \\
Cluster & Cluster & $z$ & $\sigma$ $\pm{\delta}_{\sigma}$ \\
&&& $\rm km \, s^{-1}$ \\
\hline
 Cl\,1232.5-1250     &  Cl\,1232     & 0.5414  & 1080 $_{-89}^{+119}$ \\ 
 Cl\,1216.8-1201     &  Cl\,1216     & 0.7943  & 1018 $_{-77}^{+73}$  \\ 
 Cl\,1138.2-1133     &  Cl\,1138     & 0.4796  &  732 $_{-76}^{+72}$  \\ 
 Cl\,1411.1-1148     &  Cl\,1411     & 0.5195  &  710 $_{-133}^{+125}$\\ 
 Cl\,1301.7-1139     &  Cl\,1301     & 0.4828  &  687 $_{-86}^{+81}$  \\ 
 Cl\,1354.2-1230     &  Cl\,1354     & 0.7620  &  648 $_{-110}^{+105}$ \\ 
 Cl\,1353.0-1137     &  Cl\,1353     & 0.5882  &  666 $_{-139}^{+136}$ \\ 
 Cl\,1054.4-1146     &  Cl\,1054-11  & 0.6972  &  589 $_{-70}^{+78}$  \\ 
 Cl\,1227.9-1138     &  Cl\,1227     & 0.6357  &  574 $_{-75}^{+72}$  \\ 
 Cl\,1202.7-1224     &  Cl\,1202     & 0.4240  &  518 $_{-104}^{+92}$ \\ 
 Cl\,1059.2-1253     &  Cl\,1059     & 0.4564  &  510 $_{-56}^{+52}$  \\ 
 Cl\,1054.7-1245     &  Cl\,1054-12  & 0.7498  &  504 $_{-65}^{+113}$ \\ 
 Cl\,1018.8-1211     &  Cl\,1018     & 0.4734  &  486 $_{-63}^{+59}$  \\ 
 Cl\,1040.7-1155     &  Cl\,1040     & 0.7043  &  418 $_{-46}^{+55}$  \\ 
 Cl\,1037.9-1243$^{\star}$     &  Cl\,1037     & 0.5783  &  319 $_{-52}^{+53}$  \\ 
 Cl\,1103.7-1245$^{\star}$      &  Cl\,1103     & 0.7031  &  252 $_{-85}^{+65}$ \\ 
 Cl\,1420.3-1236     &  Cl\,1420     & 0.4962  &  218 $_{-50}^{+43}$  \\ 
 Cl\,1119.3-1129     &  Cl\,1119     & 0.5500  &  166 $_{-29}^{+27}$  \\
\hline
\end{tabular}
\end{minipage}
Col. (1): Cluster name.  Col. (2): Short cluster name. 
Col. (3) Cluster redshift. Col. (4) Cluster velocity dispersion. 
Redshifts and velocity dispersions are taken from Halliday et al. (2004) and
Milvang-Jensen et al. (2008). Clusters with an asterisk do not have
local density measurements, and only their numbers of members
are used in this paper.
\end{table}

\section{Simulation}

We make use of the Millennium Simulation (hereafter MS; Springel et
al. 2005).  The MS follows $N= 2160^3$ particles of mass
$8.6\times10^{8}\,h^{-1}{\rm M}_{\odot}$ within a comoving box of size
$500\, h^{-1}$Mpc on a side and with a spatial resolution of $5\,
h^{-1}$kpc.  We extracted 90 haloes at $z=0.6$ and 90 at $z=0$ within
the simulation box, uniformly distributed in log(mass) between
$5\times10^{12}\,{\rm M}_{\odot}$ and $5\times10^{15}\,{\rm
M}_{\odot}$. These are $M_{200}$ masses (hereafter ``halo masses''),
computed from the simulations as follows.  We define $R_{200}^{MS}$
of a FOF-halo as the radius of a sphere which is centered on the most
bound particle of the group and has an overdensity of 200 with respect
to the critical density of the Universe af the redshift considered.
The enclosed mass is defined as $M_{200}$.
For 10-neighbours density measurements, only
haloes with more than 10 galaxies brighter than $M_V=-20$
were considered (65 haloes at $z=0.6$ and 62 at  $z=0$).
Dark matter haloes were populated using the
semi-analytic model presented in De Lucia \& Blaizot (2007), which
are publicly available (see also
Croton et al. 2006).
When deriving quantities to be compared with observations (see next section), 
in order to take into account projection effects of galaxies close, 
but not belonging, to the halo, 
for each halo we considered a box 6Mpc on a side, centred on the most bound
particle.

For each halo, we considered all galaxies
within 2  $R_{200}$ radii
from the central galaxy to compute the projected
velocity dispersion along the x, y, and z axes using the same bi-weight
estimator that was used for the EDisCS clusters, and used
the mean of these projected velocity dispersions as the velocity dispersion
of the system.

\section{Methods -- Observed and simulated measurements}

In this section we describe the basic ingredients of our study,
and the nomenclature adopted throughout the paper.

For simplicity, in the following we will refer to all galaxy systems
of any $\sigma$ and mass as ``clusters'', because the distinction between
clusters and groups is irrelevant in this paper.

Our analysis is performed within $R_{200}$, defined as the radius
delimiting a sphere with interior mean density 200 times the critical
density. $R_{200}$ is often assumed to be approximately equal to the
cluster virial radius, since the radius corresponding to a mean
interior overdensity of $\sim 200$ has been shown by N-body
simulations to ``accurately demarcate the virialized region of the DM
halo, which is in approximate dynamical equilibrium, from the
exterior, where material is still falling in'' (Cole \& Lacey 1996).
Since most observational studies use $R_{200}$ to compare galaxy
properties in clusters at different redshifts, or in clusters and
groups of different masses, we use $R_{200}$ to investigate
how the density felt by those galaxies that are usually compared
observationally changes with redshift and halo mass.

Assuming the virial theorem is valid, from the
evolution of the critical density with redshift, $R_{200}$ can be
estimated from the observed line-of-sight velocity dispersion
as (Finn et al. 2005):
\begin{equation}
R_{200} = 1.73 \, \frac{\sigma}{1000 \, \rm km \, s^{-1}} \,
{\frac{1}{\sqrt{{\Omega}_{\Lambda} + {\Omega}_{0}(1+z)^3}}} \, h^{-1}_{100}
\, \rm Mpc
\end{equation}

Similarly, from the virial theorem, the cluster mass can be estimated 
from the velocity dispersion as $M = 1.2 \times 10^{15}
(\frac{\sigma}{1000 \, \rm km \, s^{-1}})^3 \times
\frac{1}{\sqrt{{{\Omega}_{\lambda}}+{\Omega}_{0}(1+z)^3}} \, h^{-1}_{100} \,
M_{\odot} $ (Finn et al. 2005). In this paper, observed 
cluster masses are therefore virialized total masses based on observed
velocity dispersions derived from spectroscopy. 
For EDisCS, the latter are in rather good agreement with
the velocity dispersions obtained from the weak lensing analysis
of Clowe et al. (2006), as shown in Milvang-Jensen et al. (2008).

In the following, we refer to an output value of the simulations
either as a ``simulated'' or a ``sim-observed'' quantity:

-- a {\it sim-observed} quantity (number of galaxies, velocity
dispersion, projected local density etc.) is computed from the
simulation with the same method that would be used {\it observationally}. For
example, the sim-observed number of members within $R_{200}$ is the
number {\it in projection} on the XY plane within $R_{200}$ (derived from the
velocity dispersion using eqn.(1)) and with a velocity along the Z axis
within 3$\sigma$ from the cluster velocity. No attempt is made to
reproduce the cluster selection strategy adopted by EDisCS, nor
the geometrical constraints in constructing the masks etc.

-- a {\it simulated} quantity gives the ``true'' value provided by the
simulation. In the example above, this would be 
the actual number of galaxies inside the sphere defined by
the halo $R_{200}^{MS}$ radius.

The difference between simulated and sim-observed quantities is
fundamental, as we will show in the paper. Obviously,
only sim-observed quantities can be directly compared with observations,
but simulated quantities are very useful to address projection issues
and other possible observational biases.

\subsection{Local densities}

All densities in this paper are computed using {\it proper} (not comoving)
quantities, i.e.  proper areas and volumes, in $\rm Mpc^{2}$ and $\rm
Mpc^{3}$. 
This is motivated by the fact
that, in order to study the dependence of
galaxy properties on the density of the local environment,
what matters are
gravitational and vicinity effects, and therefore proper distances
between galaxies.

Projected local galaxy densities were computed from the observations
as described in detail
in Poggianti et al. (2008) using the circular area that in projection
on the sky encloses the 10 closest galaxies brighter than an absolute
V magnitude $M_V=-20$. 

From the simulation, we compute both the sim-observed
projected 2D local density and the physical 3D density.

Sim-observed 
projected local densities are computed for each galaxy ``member'' 
(sim-observed member)
of a simulated cluster as for observations:
we consider as members all galaxies brighter than a limit $M_V$ (either
-20 or -19.4, see below) within a
projected radius $R_{200}$ from the ``BCG'' (the central galaxy of the
halo) and within 3$\sigma$ in velocity from the mean 
cluster velocity.
The projected local density
$\Sigma$ (gal/$\rm Mpc^2$)
is calculated from the ratio $n$/Area where Area is the
circular area encompassing the $n$ projected closest neighbours
brighter than $M_V$ in the XY (``sky'') plane, with $n=10$ unless
otherwise stated. 

The ``physical'' local density $\rho$ (gal/$\rm Mpc^3$) is
meant to measure the 3D number density of a region around each galaxy. 
It is computed as the ratio $n$/Volume, where the Volume is the spherical
volume encompassing the $n$ 
closest neighbours in 3D brighter than $M_V$, with
$n=10$ unless otherwise stated. 
The 10 closest neighbours to each galaxy are those with the
10 smallest values of distance $d= sqrt({(\delta x)}^2 + {(\delta
y)}^2 + {(\delta z)}^2)$ from the galaxy, where $(\delta x, \delta y,
\delta z)$ are the distances along each axis in the simulation.  

When computing the densities from the simulation, 
all neighbours above the magnitude cut were
considered, also if outside $R_{200}^{MS}$, to avoid edge effects at the radius
$R_{200}^{MS}$, similarly to what was done observationally.

\section{Results - The evolution of the local density distribution}

\begin{figure*}
\includegraphics[width=18cm]{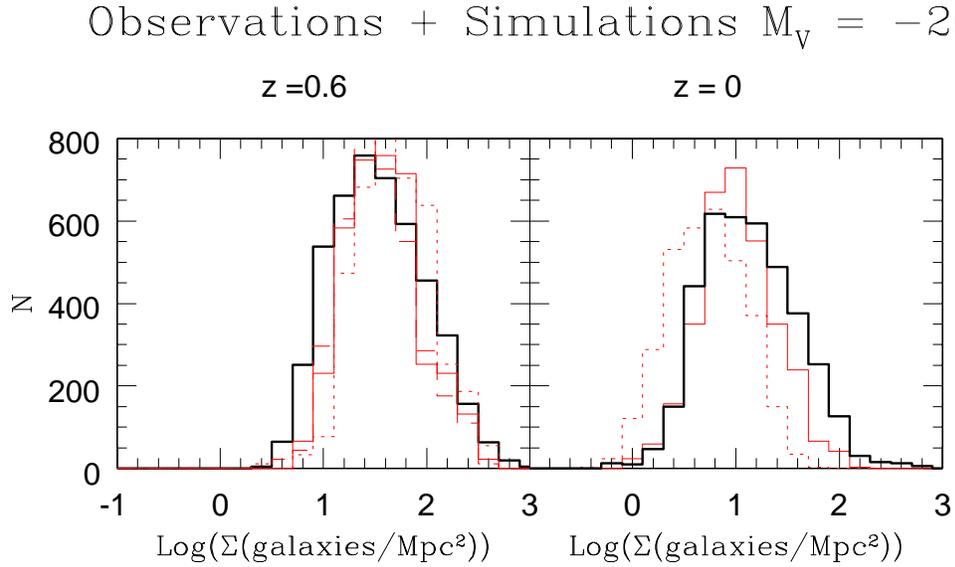}
\caption{Observed projected local density (galaxies per $\rm Mpc^{2}$)
distributions (thin histograms, in red) of spectroscopic members within the
$R_{200}$ of EDisCS clusters at $z \sim
0.6$ (left panel) and in Sloan clusters at $z=0.04-0.1$ (right panel).
They are compared, (thick histograms, in black), 
with the sim-observed distributions within the
projected $R_{200}$ in Millennium Simulation haloes of masses
M$>10^{14}$ at $z=0.6$ and $z=0$.  A fixed magnitude limit $M_V=-20$
is adopted in both cases. In the case of observations, different
density estimates are shown, as described in detail in Poggianti et
al. (2008): for EDisCS (left panel), using a statistical background
subtraction (short dashed histogram), using photo-z integrated
probabilities for membership (thin solid histogram) and the best photo-z
estimate (long dashed histogram); for Sloan (right panel), using a
statistical background subtraction (thin solid histogram) or using only the
spectroscopic catalog (dashed histogram). The latter suffers from spectroscopic
incompleteness in SDSS clusters. 
\label{uffa555}}
\end{figure*}

Observationally, the projected local density distribution in low-z
clusters is shifted towards lower values compared to clusters of
similar masses at higher redshift.  This is shown in
Fig.~\ref{uffa555}, and was mentioned in Poggianti et al. (2008).
This could also have been inferred if someone had compared previous
low-z (Dressler 1980, Lewis et al. 2002) and high-z (Dressler et
al. 1997, Postman et al. 2005) cluster observations.

To our knowledge, the evolution of the observed density distribution,
its origin and consequences have not been discussed in the past. 
It may be argued that this effect could be due to density measurements at
high-z suffering from a higher contamination of interlopers, that can
artificially inflate the observed densities, but we will show this is
not the case.

In order to understand the significance of the density evolution, we 
investigate how the observed 2D projected density distribution at
different redshifts is related to the physical 3D density distribution
in the simulation, and how both are expected to change with redshift
in clusters.

Observations and simulations are compared in Fig.~\ref{uffa555}
for galaxies brighter than $M_V \leq -20$. Only
haloes with $M>10^{14}\,{\rm M}_{\odot}$ are used in this comparison,
to match the masses of EDisCS and SDSS clusters that dominate
the observed density distribution. 

The observed projected density distribution (red histograms)
shifts by about 0.4dex from $z=0.6$ to $z=0$, 
corresponding to a decrease by a factor $2.5$.
Sim-observed projected densities evolve by the same 
amount (black histograms), and they follow rather well
the observed projected density distributions both at high- and
low-z. 

\begin{figure*}
\centerline{\hspace{1cm}\includegraphics[width=18cm]{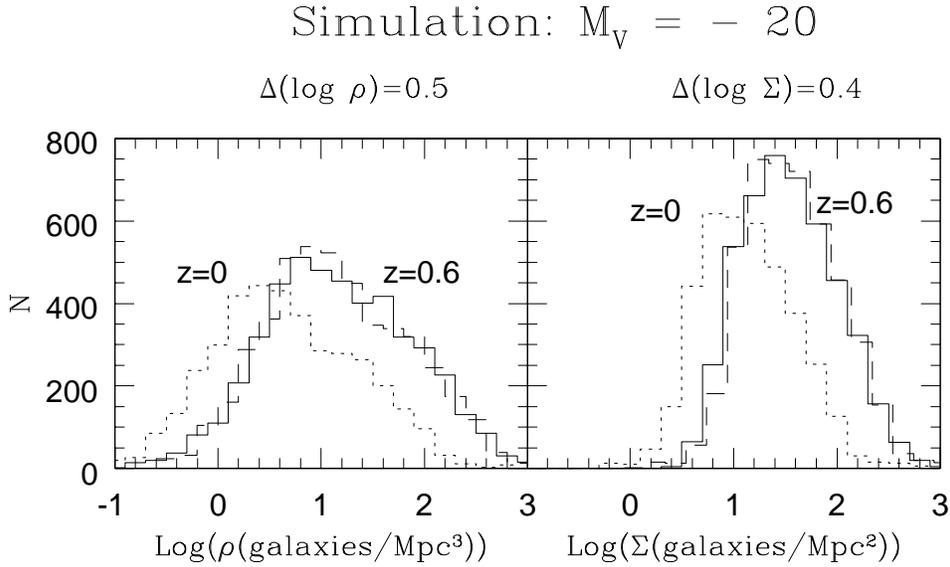}}
\caption{Physical density (left, galaxies per $\rm Mpc^{3}$) 
and sim-observed projected local density (right, galaxies per $\rm Mpc^{2}$)
distributions of galaxies within the projected $R_{200}$
in Millennium Simulation haloes of masses M$>10^{14}$ 
at $z=0.6$ (solid histograms) and $z=0$ (dotted
histograms). A fixed magnitude limit $M_V=-20$ is adopted at both
redshifts.
The right panel was compared with the observed projected local density
distributions in Fig.~\ref{uffa555}.
The long dashed histogram is the $z=0$ distribution shifted by 
0.5dex (left panel) and 0.4dex (right panel), 
normalized to the same number of galaxies of
the corresponding $z=0.6$ distribution. The dashed histogram in the right 
panel has actually been shifted by 0.44dex instead of 0.4
to prevent the histograms from completely overlapping.
\label{uffa55}}
\end{figure*}

The corresponding 3D physical densities computed from the simulation
are shown in the left panel of Fig.~\ref{uffa55}.  The evolution of
the physical 3D densities is even stronger than
that of the 2D distribution.  Physical densities for
$M_V=-20$ were on average higher by 0.5 dex, that is a factor $\sim 3.2$,
in distant clusters than today.  Therefore, the evolution
in projected density 
is not merely a projection effect with $z$, but
corresponds to physically evolving conditions in 3D galaxy
number density.

It is also interesting to note that, when accounting for the average observed
shift and rescaling to the same number of galaxies, 
the 2D and 3D density distributions at $z=0$ closely match the
respective 
distributions at $z=0.6$ (see dashed histograms in Fig.~\ref{uffa55}). 
After number rescaling,
the high-z distribution can be obtained with good approximation
from the low-z histogram
simply multiplying each density by a factor
$\sim 3.2$ in 3D, and 2.5 in 2D, since the shape of the distribution does 
not change.

\begin{figure*}
\centerline{\hspace{1cm}\includegraphics[width=18cm]{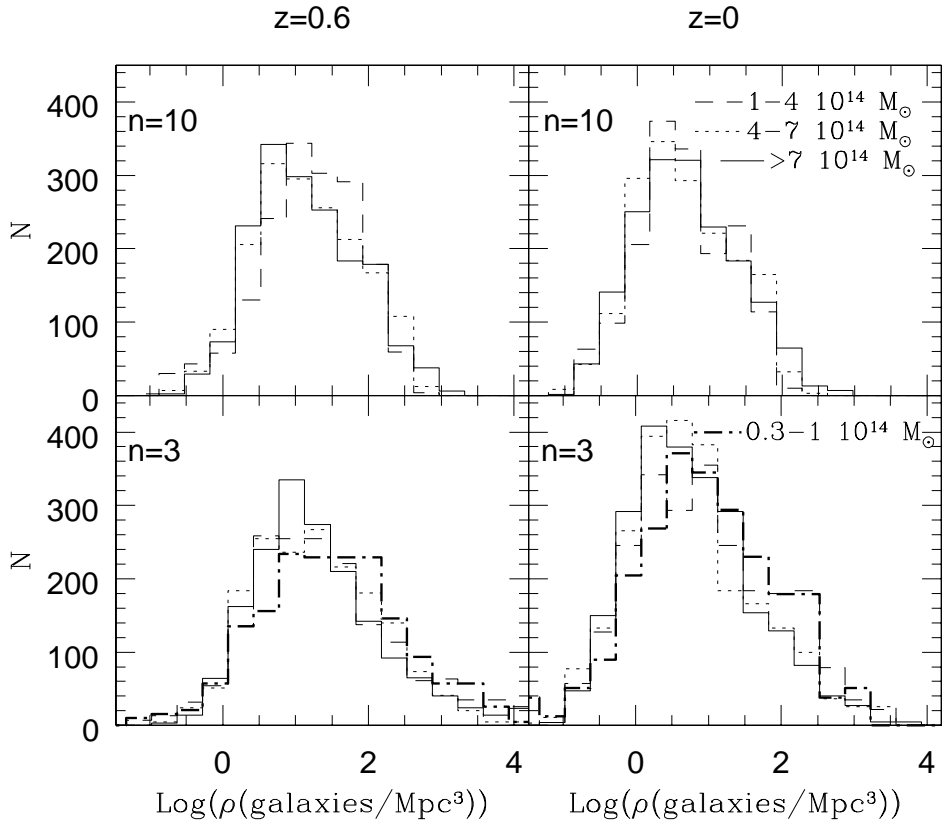}}
\caption{Physical density distribution
(galaxies per $\rm Mpc^3$) of galaxies in Millennium Simulation haloes
at $z=0.6$ (left panels) and $z=0$ (right panels) for different DM
halo mass ranges. Top and bottom panels: density computed using the 10 and 3
closest neighbours, respectively. Halo mass ranges: 
$> 7 \times 10^{14} \, M_{\odot}$ (solid histogram), 
$4-7 \times 10^{14} \, M_{\odot}$ (dotted histogram),
$1-4 \times 10^{14} \, M_{\odot}$ (dashed histogram),
$0.3-1 \times 10^{14} \, M_{\odot}$ (heavy dot-dashed histogram, only in 
bottom panels).
\label{4masses}}
\end{figure*}

Importantly, we find that both the 2D and the 3D density distributions
do not vary strongly with cluster mass, as shown in
Fig.~\ref{4masses}.\footnote{Although a wider study as a function of
the cluster delimiting radius is beyond the scope of this paper, we
have verified that the galaxy 3D number density distribution does not vary
strongly with cluster mass also when using a smaller radius,
$R_{500}$, the radius whose interior mean density is equal to 500
times the critical density.} The $n=10$ neighbours physical density
distribution are rather similar at all halo masses above
$M=10^{14}\,{\rm M}_{\odot}$ (top panels), covering the same range of
densities. The binned distributions shown in Fig.~\ref{4masses} do not
overlap with each other for always less than 12\% of the galaxies.  At
lower halo masses, $M = 0.3 - 1 \times 10^{14}\,{\rm M}_{\odot}$,
corresponding to the lowest mass groups in EDisCS, using $n=10$ is
inappropriate because the number of luminous members is less than 10
for many systems. Using densities computed for $n=3$ (bottom panels in
Fig.~\ref{4masses}), the density distributions remain rather similar
to those of more massive clusters. Galaxies in haloes with masses $M =
0.3 - 1 \times 10^{14}\,{\rm M}_{\odot}$ cover the same density range
of galaxies in more massive haloes, and the 3D density distributions
do not overlap in less than 15\% of the galaxies.  Remarkably, this
means that the distributions of local densities to the 3 closest
neighbours are quite similar for galaxies in clusters over two orders
of magnitude in mass, suggesting that there is an approximately equal
distribution of dense and less dense regions from groups to massive
clusters when the ``really local'' density is considered.  The
intuitive belief that more massive clusters are denser environments
proves to be wrong. Actually, we see that the lowest halo mass bin has
the largest 3D high-density tail.

So far we have used a fixed $M_V=-20$ galaxy magnitude limit at both
redshifts, but it is physically more meaningful to adopt an $M_V$
limit evolving with redshift, taking into account the fact that a
galaxy luminosity evolves as the galaxy stars become older. We
consider passive evolution, that is the evolution of an integrated
spectrum due to simple aging of old stellar populations, without the
addition of any new star.  Using van Dokkum \& Franx (2001) or,
equivalently, Fritz et al. (2007) spectrophometric models for a single
episode of star formation occurred at $z$ between 1.5 and 3 for a
Salpeter IMF and solar metallicity, the evolution of $M_V$ between
$z=0.6$ and $0$ is 0.5-0.7mag.\footnote{As pointed out by Conroy, Gunn
\& White (2009), the uncertainties in the IMF slope at low stellar
masses translate into an uncertainty in the passively evolving
luminosity evolution of 0.4mag in $K$ between $z=1$ and $z=0$.  If we
arbitrarily assume the uncertainty to be the same in $V$ and in $K$,
the change in luminosity over our redshift range due to the IMF
is still much smaller than that obtained
lowering the star formation redshift to e.g. $z=1$. Thus, the
uncertainty in the passively evolving magnitude limit is dominated by the
assumption in the formation redshift of most stars. The great majority
of cluster galaxies have very old mass-weighted stellar ages
(e.g. Fritz et al. 2010, in prep.) therefore a high formation redshift
range as the one adopted here is appropriate.} Therefore, we consider an
evolving magnitude limit going from $M_V=-20$ at $z=0.6$ to
$M_V=-19.4$ at $z=0$.

Passive evolution produces a stronger
decreasing luminosity evolution than any other star formation
history. Therefore, assuming a passively evolving magnitude limit, we
are sure to include in the density calculation at $z=0$ all galaxies
that would be included at $z=0.6$, and possibly some more galaxies
that do not make it into the sample at $z=0.6$.  This yields a sort of
lower limit on (but is probably close to)\footnote{It should be close
to the evolution of the density distribution in a galaxy stellar mass
limited sample because the majority of bright galaxies as those
considered here have a negative luminosity evolution very close to
passive, having truly passively evolving or declining star formation
histories.}  the evolution of the density distribution as it would be
measured using a fixed galaxy stellar mass limit and if galaxy masses
did not evolve significantly.\footnote{Note that the simulation, by
its own nature, does not assume conservation of galaxy numbers, nor
galaxy masses, since it includes galaxy mergers.}  On the contrary,
the non-evolving $M_V=-20$ used above excludes from the low-z density
calculation all those galaxies that have faded below the limit by
$z=0$ but were included at $z=0.6$. Therefore, it provides an upper
limit for the evolution of the density distribution for a mass-limited
galaxy sample with non-evolving galaxy masses.

Unfortunately, 
the SDSS spectroscopy is not deep enough 
to reach $M_V=-19.4$ in our SDSS clusters. Adopting a brighter, 
passively evolving limit at both redshifts would result in too few
galaxies in high-z clusters to allow a robust density determination.
Hence, we are forced to investigate the passive evolution case only in
the simulation, mimicking the way a sufficiently deep observational survey
at low-z would be treated to compare with the high-z
results. The fact that the observed
$M_V = -20$ density distributions are well reproduced in the
simulation, as is the number of cluster members shown in the next
section, makes us confident that the simulated densities should be
reliable also down to 0.5 magnitude fainter.

\begin{figure*}
\centerline{\hspace{1cm}\includegraphics[width=18cm]{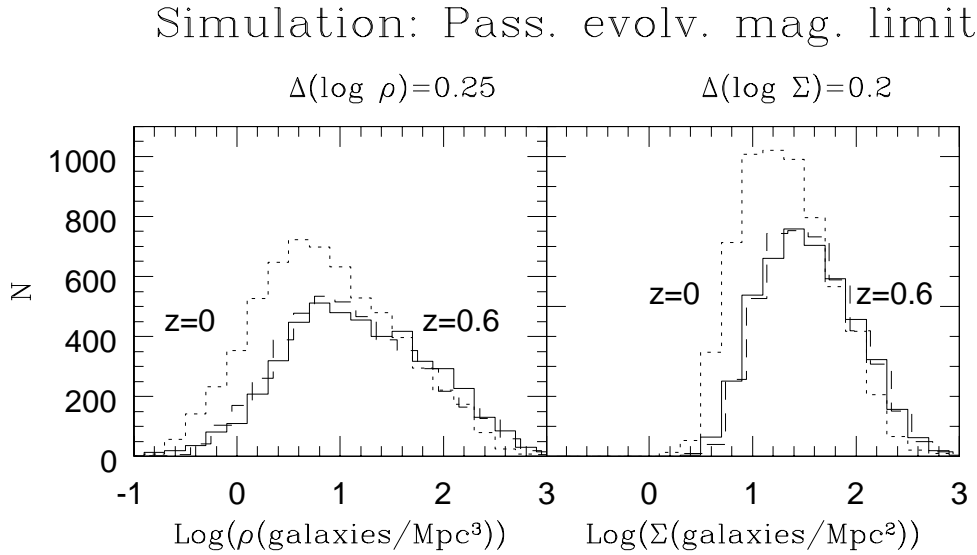}}
\caption{As Fig.~\ref{uffa55} but using a passively evolving galaxy
magnitude limit: $M_V=-20$ at $z=0.6$ and $M_V=-19.4$ at $z=0$. 
Physical density (left, galaxies per $\rm Mpc^{3}$) 
and sim-observed projected local density (right, galaxies per $\rm Mpc^{2}$)
distributions of galaxies within the projected $R_{200}$
in Millennium Simulation haloes of masses M$>10^{14}$ 
at $z=0.6$ (filled histograms) and $z=0$ (dotted
histograms). 
The long dashed histogram is the $z=0$ distribution shifted by 0.25dex 
(left panel)
or 0.2dex (right panel) normalized to the same number of galaxies as
the corresponding $z=0.6$ distribution. The dashed histogram in the right 
panel has actually been shifted by 0.24dex instead of 0.2 
to prevent the histograms
from completely overlapping.
\label{uffa55bis}}
\end{figure*}

With a passively evolving
galaxy magnitude
limit, 
the shifts of the 2D and 3D density distributions in the simulation
are smaller, but still
conspicuous (Fig.~\ref{uffa55bis}). After number rescaling,
the high-z 3D and 2D density
distributions match the $z=0$ distributions 
if densities
decrease by a factor $\sim 1.8$ (0.25dex) in 3D and 1.6 (0.2dex) in 2D.

The analysis of the simulations presented so far demonstrates that,
within $R_{200}$ and computing local densities 
from the 10 or 3 closest neighbours,
 a) distant clusters were denser
environments than low-z clusters, in the sense that, at $z=0.6$,
cluster galaxies had on average $\sim 2 - 3$ times ($>0.25$dex, 
$<0.5$dex) more luminous/massive neighbors than at $z=0$, and 
b) the distribution of physical densities is rather similar in
haloes over two orders of magnitude in mass.

\section{The cause of the density evolution in a cosmological framework}

In the analysis presented above, densities represent {\it number
densities} of galaxies, that means number of neighbours of an
individual galaxy down to a certain magnitude, per unit of projected
area or volume. In the following we want to establish

\begin{itemize}
\item
how the evolution in number density of individual galaxies corresponds
to the evolution of the {\it global average number density} over the
whole cluster (total number of members per unit volume $N/V$).  In
doing this, we investigate separately the evolution of $N$ within
$R_{200}$ and the evolution of the volume $V$ enclosed by $R_{200}$.
The question we wish to address is how the change of the 3D
number density with redshift is related 
to the evolution of $N/V$, and $N$
and $V$ separately, and how and why these evolve. Note that $N/V$ can
be estimated observationally,
by estimating a
cluster radius from either $\sigma$ or an alternative mass measurement,
and counting the number of members within this radius.

\item how the {\it number} density evolution
is related to the mean {\it matter} density evolution
in clusters, that is the evolution of the mean total, or DM,
mass per unit volume.
In principle, mass density and number density are different measurements of 
environment, and are expected to influence physical processes
in different ways: for example, 
number density is more relevant for galaxy-galaxy 
interactions and mergers, while mass density is more relevant for
other effects, such as the cluster tidal field.
We wish to understand if number densities and mass densities evolve
in similar or different ways, to gain a more complete picture
of the evolution of environmental conditions.
\end{itemize}

\begin{figure*}
\centerline{\includegraphics[width=16cm]{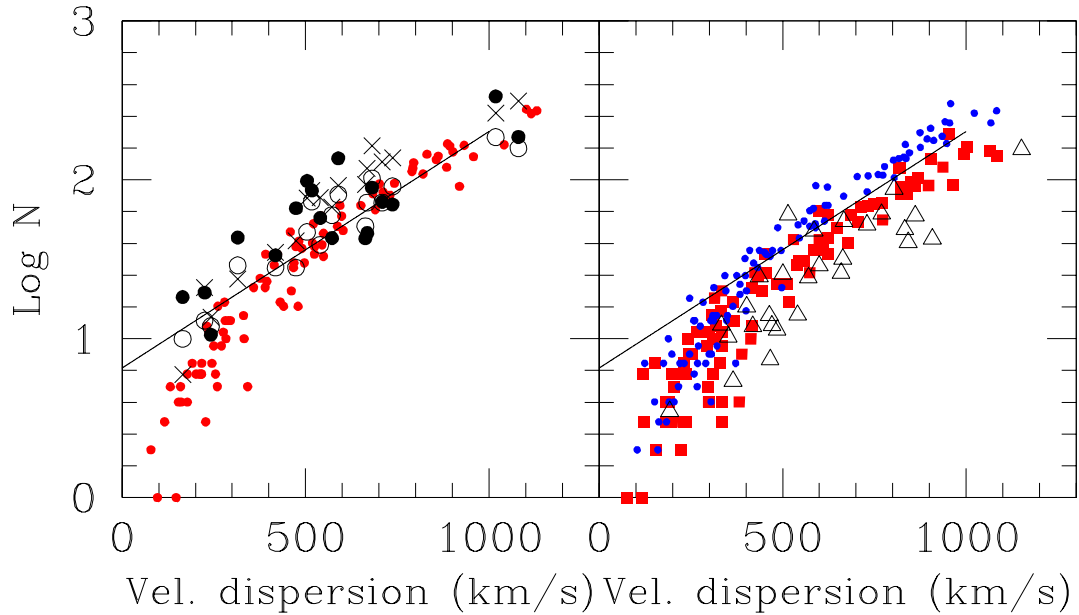}}
\caption{Number of members within $R_{200}$ versus cluster velocity
dispersion.  {\bf Left: high-z ($z\sim 0.6$).} EDisCS datapoints are black
circles (computed using the statistical subtraction), empty circles
(computed from photo-z membership) and crosses (computed as the number
of spectroscopic members corrected for incompleteness), see text for details.
Theoretical (MS) sim-observed
values at $z=0.6$ are represented as small red circles. The
solid line is the best linear fit to the theoretical values at $\sigma > 400
\, \rm km \, s^{-1}$. In all cases only galaxies with $M_V \leq -20.0$ are 
included. {\bf Right: low-z ($z\sim 0.07$).} Sloan datapoints
at $z=0.04-0.1$ for $M_V \leq -20.0$, computed as the number of
spectroscopic members corrected for incompleteness, 
are empty triangles. Theoretical
(MS) sim-observed values at $z=0.0$ for $M_V \leq -20.0$ are represented as red
squares. The solid line is the best fit to the $z=0.6$ $M_V \leq
-20.0$ theoretical values, repeated from the left panel. Blue small
points are theoretical values at $z=0$ adopting a passively evolving
galaxy magnitude limit ($M_V \leq -19.4$).}
\label{vssigma}
 \end{figure*}

The number of galaxies per cluster observed in the EDisCS and Sloan
datasets within $R_{200}$
is shown as a function of the observed velocity dispersion in
Fig.~\ref{vssigma}. This is compared at each redshift with the
sim-observed number of galaxies as a function of sim-observed
velocity dispersion. From Fig.~\ref{vssigma}, it can be seen that
the sim-observed numbers agree quite well with
the observed numbers at both redshifts, although the sim-observed numbers
at $z=0$ tend to be slightly higher than the observed values by about 0.1dex.

The number of galaxies increases with velocity dispersion at both
redshifts (higher velocity dispersion clusters have more members,
obviously). For a fixed $M_V \leq -20$ limit, the number of members decreases
from $z \sim 0.6$ to $z \sim 0$ (the number of galaxies down to a
fixed magnitude limit in a cluster of a given $\sigma$ was 
higher in the past), 
as shown by the offset in the right
panel between the high-z best-fit line and the $z=0$ points.

\begin{figure*}
\centerline{\includegraphics[width=10cm]{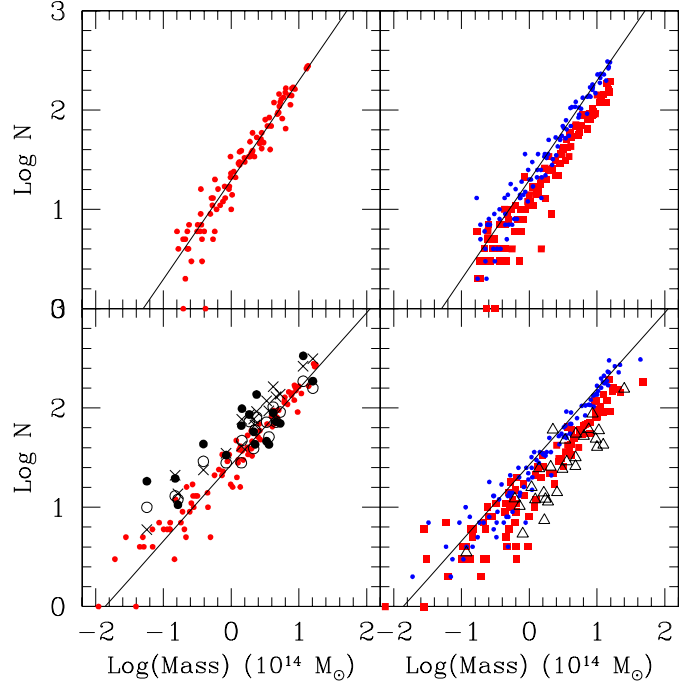}}
\caption{{\bf Bottom left and right} 
Number of members within $R_{200}$ versus cluster mass. The cluster
mass is computed from the observed or sim-observed velocity dispersion.
{{\bf Left: high-z} ($z\sim 0.6$).} EDisCS datapoints are black
circles (computed using the statistical subtraction), empty circles
(computed from photo-z membership) and crosses (computed as the number
of spectroscopic members corrected for incompleteness).
Theoretical (MS) sim-observed
values at $z=0.6$ are represented as small red circles. The
solid line is the best linear fit to the theoretical values at $\sigma > 400
\, \rm km \, s^{-1}$, $LogN = (0.77\pm0.03) \times LogM + (1.43\pm0.02)$. 
In all cases only galaxies with $M_V \leq -20.0$ are 
included. {{\bf Right: low-z} ($z\sim 0.07$).} Sloan datapoints
at $z=0.04-0.1$ for $M_V \leq -20.0$, computed as the number of
spectroscopic members corrected for incompleteness, 
are empty triangles. Theoretical
(MS) sim-observed values at $z=0.0$ for $M_V \leq -20.0$ are represented as red
squares. The solid line is the best fit to the $z=0.6$ $M_V \leq
-20.0$ theoretical values, repeated from the left panel. Blue small
points are theoretical values at $z=0$ adopting a passively evolving
galaxy magnitude limit ($M_V \leq -19.4$.)
{\bf Top left and right} Only simulations are shown, with same symbols as
in the bottom panels. The cluster mass plotted is now the
dark matter halo mass. The relation steepens, and the slope is now 1:
$LogN = (1.00\pm0.04) \times LogM + (1.29\pm0.02)$.
}
\label{vsmass}
 \end{figure*}

Since clusters with the same observed velocity dispersion but different
redshift have different masses  (cf. \S4), 
we remove this effect by plotting the number of
galaxies versus cluster mass in Fig.~\ref{vsmass}, 
see bottom panels. 
The trends remain similar: both observationally and theoretically, the
number of galaxies increases linearly with mass in a log-log plot, 
and, for a fixed galaxy magnitude limit, it decreases towards lower redshift.

To understand if the number decline at lower redshift is simply due to
galaxies dropping off the sample at $z=0$ because they fade below the
fixed magnitude cut, in Fig.~\ref{vssigma} and Fig.~\ref{vsmass} we
also show the sim-observed number of cluster members for a passively
evolving magnitude limit $M_V = -19.4$ at $z=0$ (blue points). 
As explained in \S5, we cannot do the same observationally, because
the SDSS spectroscopy is not deep enough. With
the passively evolving limit, the numbers at $z=0$ in the bottom
right panel of
Fig.~\ref{vsmass} are only 0.057dex lower than the best fit $z=0.6$
correlation shown by the solid line ($LogN = (0.77\pm0.03) \times LogM
+ (1.43\pm0.02$)). As we will discuss later in this section, this
residual small mismatch is caused by the way masses are derived, and
the effect disappears when ``true'' simulated masses of haloes are
considered (top right panel in Fig.~\ref{vsmass}).

To conclude, both observations and simulations show a mild
evolution in the $N$ vs $M$ relation with redshift for a fixed
magnitude limit.
When passive evolution is taken into account, the number
of members in a cluster of a given mass 
does not evolve between $z\sim 0.6$ and $z=0$.  A similar result was
found by Lin et al. (2006), who, based on cluster mass estimates
derived from the X-ray temperature and observed number counts,
concluded that for passive evolution the $N-M$ relation shows no sign
of evolution out to $z=0.9$.

This result implies that the evolution in the number
density distributions for the passive evolution case
in cluster samples with the
same cluster mass distribution at different redshifts 
(Fig.~\ref{uffa55bis}) is not due to a
higher number of members at $z=0.6$, but must originate mostly from
the ``expanded size'' of local clusters compared to distant ones, that
is, to the evolution of the cluster volume.

\subsubsection{A recollection of the evolution of cluster volume
and dark matter density}

At this point it is useful to recall some textbook knowledge
about how and why the volume and the mean
matter density of clusters evolve, and how
this evolution simply stems
from using $R_{200}$ to define both of them. 

Due to the expansion of the universe, a unit volume today corresponds
to a volume smaller by a factor $1/(1+z)^3$ at redshift $z$ (Peebles 1993). 
Therefore,
the physical 3D density (number of galaxies per unit volume) of
non-evolving objects locked into the Hubble flow changes with redshift
as $(1+z)^3$. 
Galaxies and clusters of galaxies, however, are far from
being unevolving objects, with clusters accreting new galaxies and
groups.
Moreover, collapsed structures such as clusters should
have broken off from the Hubble flow, at a time when their density
exceeded a fixed multiple of the critical density at that redshift 
(Peebles 1993, Cole \& Lacey 1996).

Since we only include galaxies
within the cluster $R_{200}$, {\it by definition} the
mean {\it matter} density within this radius is 200 times the critical
density at that redshift. The critical density evolves with $z$ as:

\begin{equation}
{\varrho}_c (z) = \frac{3{H_0}^2}{8\pi G} \times ({\Omega}_{\lambda} + {\Omega}_{0}(1+z)^3)
\end{equation}

where $G$ is the gravitational constant $=4.29 \times 10^{-9} \rm \, (km/s)^2 \, Mpc \, M_{\odot}$.

As a consequence,
the mean matter density (= mass per unit volume) in clusters
of {\it any mass}, within a radius $R=R_{200}$ 
goes as:

\begin{eqnarray}
{\varrho}_m (z) & = & \frac{M}{V} = \frac{M}{(4/3) \pi R_{200}^3} = 200 {\varrho}_c(z) \notag \\ & = & 200 \times \frac{3{H_0}^2}{8\pi G} \times ({\Omega}_{\lambda} + {\Omega}_{0}(1+z)^3) \notag \\
         & =  & a + b (1+z)^3 
\end{eqnarray}

where
$a= 200 \times \frac{3{H_0}^2}{8\pi G} \times {\Omega}_{\lambda} = 22.241 \times 10^{12} \, \rm Mpc^{-3} \, M_{\odot}$
and 
$b = 200 \times \frac{3{H_0}^2}{8\pi G} \times {\Omega}_{0} = 7.414 \times 10^{12} \, \rm Mpc^{-3} \, M_{\odot} $ in our cosmology.

This means that {\it the mean matter density in clusters at a 
given redshift is the same for all clusters}.
Moreover, it means that {\it the mean matter density
evolves in the same way for clusters of all masses},
by a factor ${\varrho}_1/{\varrho}_2 = (a+b(1+z_1)^3)/(a+b(1+z_2)^3)$ 
between $z_1$ and $z_2$. 
This has a number of implications, all consequences
of using $R_{200}$ as the radius delimiting the cluster,
among which:

\begin{itemize}
\item
The mean matter density within a cluster $R_{200}$ at
any redshift can be computed from eqn.(3): the most distant cluster at $z_1$
has a density $(a+b(1+z_1)^3)/(a+b(1+z_2)^3)$ times higher than a cluster
at $z_2$. 
In our case, any cluster at $z=0.6$ (of any mass) has a matter density
$52.61 \times 10^{12}\rm M_{\odot} \, \rm Mpc^{-3}$, and any cluster
at $z=0$ has ${\varrho}_m = 29.65 \times 10^{12}\rm M_{\odot} \, \rm
Mpc^{-3}$, a factor of 1.774 less dense. 
The larger the redshift difference, the stronger the evolution:
between $z=1.0(1.5)$ and $z=0$, the matter
density evolves by a factor 2.750(4.656).

\item
Equation (3) implies that
two clusters of the same mass $M_1=M_2$ but
different redshifts $z_1$ and $z_2$ have different volumes: the most
distant cluster at $z_1$ has a volume $(a+b(1+z_2)^3)/(a+b(1+z_1)^3)$
times smaller than the other (1.774 times smaller at $z=0.6$ compared to 
$z=0$). 
The ratio of the volumes of two equal mass clusters at
different redshifts is therefore invariant with the cluster mass, and
only depends on redshift, while, 
at any given redshift, the volume is
linearly proportional to the mass (eqn.~3): a cluster twice as massive has
twice as large a volume.\footnote{One might wonder how 
mass $M$ and volume $V$, separately, change in an
evolving individual cluster(halo), considering that on average $M$
grows with $z$ by a factor that is larger at larger
masses, as shown by numerous theoretical works
(eg. Lacey \& Cole 1993). 
Simulations show that the average mass growth between $z=0.6$
and $z=0$ is about 1.5
times for $1.1 \times 10^{12}$ (mass at $z=0.6$), about 1.9 for $5.6
\times 10^{13}$, 2.35 for $3.7 \times 10^{14}$, and 3.0 for $\sim 10^{15}$
(the mass growth for finer intervals of cluster masses can be found
in Table~4 of Poggianti et al. 2006). 
The independence of ${\varrho}_m$ from the cluster mass (eqn. 3)
implies that, although the mass growth rate depends on cluster
mass, the parallel evolution in cluster volume (radius) 
compensates the mass dependence.}

\end{itemize}

\subsection{The constancy of the number of galaxies per unit cluster mass,
and the evolution of the number of galaxies per unit volume}

\begin{figure*}
\centerline{\includegraphics[width=10cm]{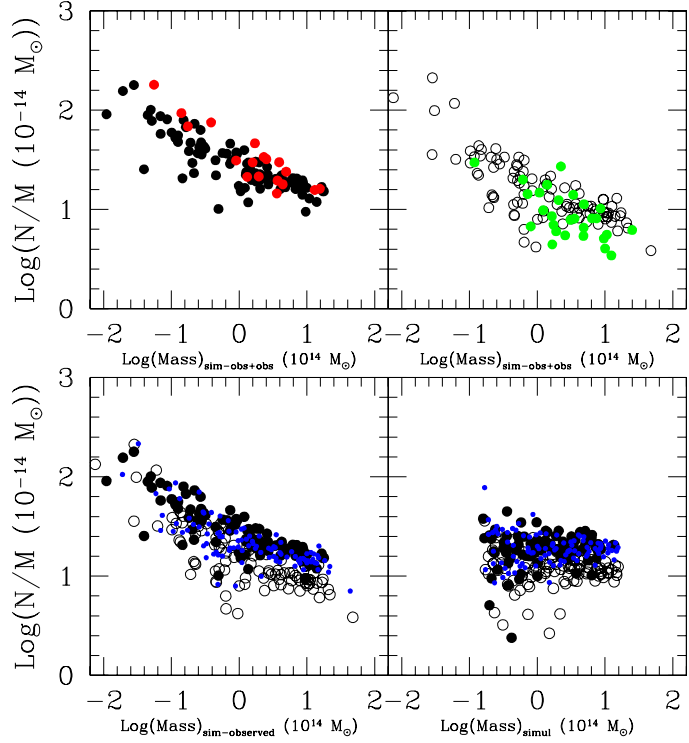}}
\caption{Number of galaxies per unit of cluster mass 
versus cluster mass. Filled black circles are from simulations
at $z=0.6$ and empty
circles are from simulations at $z=0$, both for a galaxy magnitude limit
$M_V=-20$. Blue small dots are the $z=0$ simulations for $M_V=-19.4$. 
Observed points for $M_V=-20$
are red (EDisCS) and green (SDSS) points in the top panels.
{\bf Bottom left.}  Masses, both on the x and y axes, are derived from the
sim-observed velocity dispersion. The number of galaxies is the 
sim-observed number of members. 
{\bf Bottom right.} Masses, both on the x
and y axes, are ``true'' simulated masses of dark matter haloes.
The number of galaxies is the 
sim-observed number of members. 
$Log(N/M(10^{-14}M_{\odot})) = (1.29\pm0.097)$ for haloes with 
$M>10^{14}M_{\odot}$, and
$Log(N/M(10^{-14}M_{\odot})) = (1.26\pm0.217)$ for 
$M<10^{14}M_{\odot}$, computed using together the high-z
and the low-z numbers corrected for passive evolution.
Results remain unchanged if we use
the ``real'' simulated number of members.
{\bf Top left.} The sim-observed results at $z=0.6$ (filled circles,
repeated from the bottom left panel) are compared with the
EDisCS observed points (number of spectroscopic members corrected
for incompleteness, red circles). Both are for $M_V=-20$.
{\bf Top right.} The sim-observed results at $z=0$ (empty circles,
repeated from the bottom left panel) are compared with the
Sloan observed points (number of spectroscopic members corrected
for incompleteness, green circles). Both are for $M_V=-20$.
}
\label{noverm}
\end{figure*}

The number of cluster members divided by the cluster mass $N/M$ is presented
in Fig.~\ref{noverm} as a function of cluster mass.  The sim-observed
quantities agree well with the EDisCS and SDSS datapoints 
(top left and right panels).  Both observations and simulation show a strong
trend of declining N/M with mass,
with a shift to lower N/M 
at lower $z$ when the $M_V=-20$
limit is used.  Accounting for passive evolution (blue points in the 
bottom left panel), 
the $z=0$ trend almost overlaps with that at $z=0.6$, as was already
shown in Fig.~\ref{vsmass}.

\begin{figure*}
\vspace{-10cm}
\centerline{\includegraphics[width=18cm]{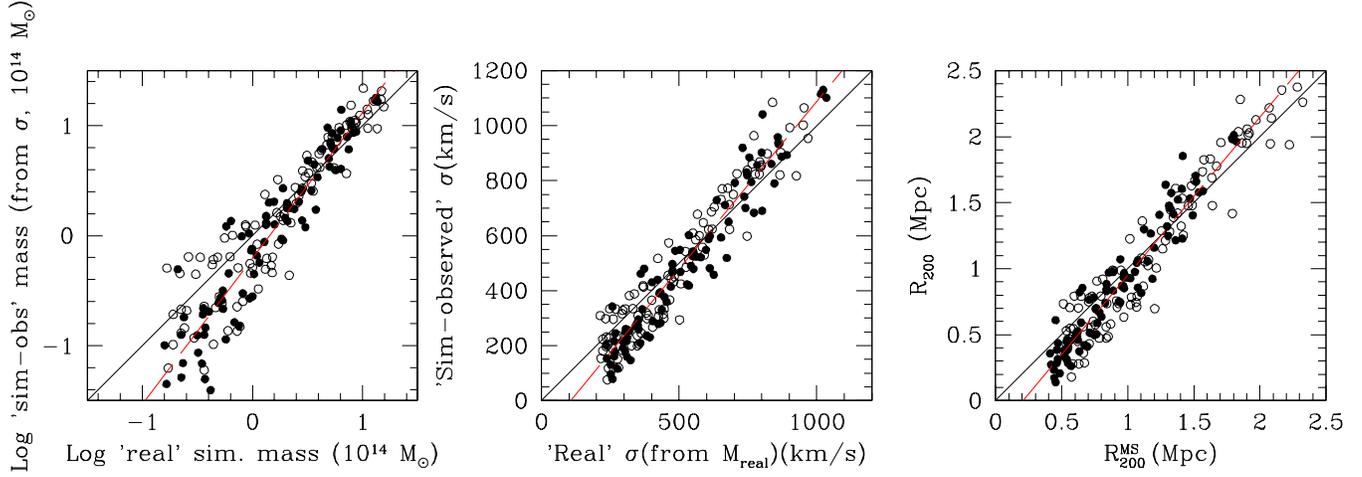}}
\vspace{-1cm}
\caption{Filled circles are simulation results at $z=0.6$, and empty
circles are at $z=0$.  {\bf Left} The ``true'' masses of Millennium
Simulation haloes are compared with the sim-observed masses of the
same haloes computed from the velocity dispersion as it would be
observed. The red dashed line is the least square fit
$log(M_{sim-obs}) = (1.325\pm0.04) \times log(M_{simul}) - (0.21\pm0.02)$.  The
solid line is the 1:1 relation.  {\bf Center} The sim-observed
velocity dispersion versus the velocity dispersion derived from the
``true'' halo mass using the M-$\sigma$ relation given in \S5.1.  The
red dashed line is the least square fit ${\sigma}_{sim-obs} = (1.22\pm0.03)
\times {\sigma}_{simul} - (129\pm16) $. The solid line is the 1:1
relation. {\bf Right} The sim-observed $R_{200}$ measured from the
simu-observed velocity dispersion using eqn.~(1) versus the halo theoretical
$R_{200}^{MS}$ radius.
The red dashed line is the least square fit ${R}_{200} = (1.20\pm0.03)
\times {R}_{200}^{MS} - (0.25\pm0.03) $. The solid line is the 1:1
relation.
\label{massmass}
}
 \end{figure*}

The observed and predicted declining $N/M$ trend with mass seems to
suggest that more massive clusters have a lower number of galaxies per
unit of cluster mass.  We will now show that this is just an illusory
result due to the way the mass is computed ``observationally''. In
fact, we find there is a discrepancy in the simulations
between the halo masses and the mass
derived from the sim-observed velocity dispersion
(Fig.~\ref{massmass}).  Masses derived from the sim-observed velocity
dispersion underestimate the halo masses by up to an order of
magnitude in low-mass groups ($M_{sim-obs} \leq 4 \times 10^{13} \,
M_{\odot}$). They yield halo mass values with good approximation
at intermediate masses ($M_{sim-obs} = 4 \times 10^{13} \, M_{\odot} -
4 \times 10^{14} \, M_{\odot}$), and slightly overestimate the halo
masses for massive clusters ($M_{sim-obs} > 4 \times 10^{14} \,
M_{\odot}$).  The correlation is $log(M_{sim-obs}) = (1.325\pm0.04)
\times M_{simul} - (0.21\pm0.02)$, where both masses are in units of
$10^{14} M_{\odot}$. The top panels of Fig.~\ref{vsmass} shows that,
using halo masses, the correlation between number of galaxies and
cluster mass steepens ($LogN = (1.00\pm0.04) \times LogM +
(1.29\pm0.02$)), and passively evolved points at $z=0$ follow the same
exact $N-M$ correlation found at $z=0.6$ with no offset (top right
panel, blue points).  

The mass discrepancy corresponds to a
discrepancy in both $\sigma$ and cluster radius: the sim-observed
velocity dispersion and sim-observed $R_{200}$ deviate from the
velocity dispersion obtained from the halo mass and from the
theoretical $R_{200}^{MS}$ radius, underestimating them at low values
and overestimating them at high values (middle and right panels of
Fig.~\ref{massmass}).

The mismatch between halo masses and cluster mass estimators based on
velocity dispersion was studied in detail by Biviano et al. (2006),
whose results agree very well with ours. 
Based on cosmological hydrodynamical simulations,
they studied clusters with masses above $10^{14} \, M_{\odot}$
considering several observational effects such as the presence of interlopers
and subclusters, sample size, incompleteness and different tracers
of gravitational potential (different types of galaxies and DM particles).
For masses below
$10^{14} \, M_{\odot}$, where we find the strongest mismatch between
the two types of masses,
we are not aware of any previous work we can compare with.

The bottom right panel
of Fig.~\ref{noverm} presents the number of cluster members per unit
cluster mass obtained using the halo
mass in the simulation instead of the sim-observed mass derived
from the sim-observed velocity dispersion. 
The decline with cluster mass disappears: the ratio
between number of members and cluster mass is constant with mass and,
when passive evolution is included, is constant 
also with redshift. There appear to
be, on average, approximately 20 galaxies brighter than a passively
evolving $z=0$ limit $M_V=-19.4$ for each $10^{14} \, M_{\odot}$ of
cluster mass, for any mass, at both redshifts.\footnote{This 
result is unchanged
if we use the ``real'' simulated instead of  the sim-observed
number of members, i.e. also the simulated average N/M ratio is constant
with mass and redshift.} 
This constancy is remarkable, also because
it applies over a two order magnitude range of cluster mass, and
regardless of redshift at least up to $z=0.6$. 

The fact that simulations are able to reproduce the observed
$Log(N/M)$ vs $LogM$ trends (upper panels of Fig.~\ref{noverm}), and
that such trends disappear when adopting the halo mass instead of the
sim-observed mass (hence, when simply removing projection effects that bias
the observational estimate of the halo mass) supports the validity of
the constancy of $N/M$, regardless of the prescriptions of the
semi-analytic model.

This constancy, and the small scatter around the mean value, especially
in massive clusters ($Log(N/M(10^{-14}M_{\odot})) = (1.29\pm0.10)$ for
haloes with $M>10^{14}M_{\odot}$, and $Log(N/M(10^{-14}M_{\odot})) =
(1.26\pm0.22)$ for $M<10^{14}M_{\odot}$), suggests that measuring $N$
should be a powerful method to estimate the cluster mass, free from
the systematic mismatch between ``true'' halo mass and $\sigma$-based
mass estimate. Recall that $N$ can be observationally estimated as
the number of members brighter than our passively evolving limit
within $R_{200}$, if an estimate of $R_{200}$ is available. The
constancy of $N/M$ should allow to start from a first guess for $N$
within a fixed metric radius (see also Andreon \& Hurn 2009), 
derive the mass, the corresponding
$R_{200}$ and $N$ within this radius, iterating the procedure until
convergence at $N/M \sim 20$ is reached. 
Further investigation,
combining simulations and observations, would be valuable to assess
the utility of this method, comparing other observed cluster samples, at
different redshifts, with simulations as in
Fig.~\ref{noverm}). This should test the model capability to
reproduce the observed $N$ 
and, most importantly, compare the mass
estimate precision with that of other methods, especially those based
on other definitions of richness that do not consider the $N/M$
constancy found in this paper
(see e.g.  Popesso et al. 2005, Gladders et al. 2007, Rozo
et al. 2009, Andreon \& Hurn 2009).

Since our simulations show that
for passive evolution and using halo masses, the average 
$N/M$ is constant both with $z$ and $M$ (Fig.~\ref{noverm}), 
and given that $V=M/(a+b(1+z)^3)$ (from eqn(3)),
the number of galaxies per unit volume goes as:

\begin{equation}
N/V = c \times (a+b(1+z)^3)
\end{equation}

where $c=N/M \sim 20 \, \rm gal/10^{14} M_{\odot}$.

Therefore, {\it both the mean matter density} (mass per unit volume,
eqn.~3) and the average {\it global number density} (number of members 
per unit volume, eqn.~4) depend only on redshift, not on cluster mass,
and are higher in more distant clusters
by the same factor $(a+b(1+z_1)^3)/(a+b(1+z_2)^3)$ with $z_1>z_2$.

We have seen that between $z=0.6$ and $z=0$ this factor is equal to
1.774 (0.25dex) in our cosmology. This is in striking agreement with,
and accounts for, the 0.25dex shift of the 3D individual number
density distribution shown in Fig.~\ref{uffa55bis}, and it must be ultimately
responsible for the higher projected number densities observed at
higher redshift (Fig.~\ref{uffa555}).

To summarize, our results show that higher-z clusters are denser than
lower-z clusters, both in individual and global number density of
galaxies, and in mass. 
The constancy of the average {\it mass} density with cluster mass, and
its evolution with redshift, is simply a consequence of the evolution of the 
critical density and of considering as ``cluster'' only the region
within $R_{200}$ which approximates the sphere of influence of
cluster gravity against the Hubble flow.

Not at all obvious was the fact that also the {\it global number} density
should follow the same laws, should be independent of cluster mass and
should evolve as the mass density. We have demonstrated that this is
true because the number of galaxies per unit of cluster mass $N/M$ is
invariant {\it both with cluster mass and redshift}, when passive
evolution is taken into account and when masses are unaffected by
observational biases. Far from obvious was also the  
evolution of the projected and the 3D galaxy number density
distributions, whose shape does not change with redshift and
whose mean evolves exactly as the global number density.

Interestingly, these conclusions apply to any virialized system
regardless of its mass, thus should be applicable from the most
massive clusters to the least massive groups. It may appear
counterintuitive, but we have just shown
that the {\it average} surrounding number and mass density seen by a galaxy
in any virialized region is the same at a given redshift, regardless
whether the galaxy belong to a small group, an intermediate mass or a
massive cluster. In the previous section, we have also shown
that not only the average density, but also
the number density {\it distribution} is approximately
the same in the virialized regions of all halo masses explored.
As far as densities are concerned, on
large and unbiased cluster samples, the mass of the system
does not matter, epoch does.

However, it is fundamental to keep in mind that, while the density average
and, approximately, distributions
are independent of cluster mass, 
there is a scatter in the predicted N/M (and therefore
N/V and 3D distribution) {\it at any given mass}, 
a scatter that increases towards lower halo masses, as visible in
Fig.~\ref{noverm} and testified by the larger scatter (0.22 versus 0.10)
in the $Log(N/M)$ relations given earlier in this section. 
Cluster-to-cluster density variations at a given
mass are therefore not described in this formalism, but should be
important for individual clusters expecially below
$10^{14} \, M_{\odot}$, and likely will depend on the specific
cluster growth history.

\section{Summary}

1) The observed distribution of projected local densities in high-z
clusters is shifted to higher values compared to the low-z
distribution, and is reproduced by simulations.  Based on the 3D
density distributions obtained from the simulations, we find that this
is due to high-z clusters being denser, in physical space, than their
local counterparts. Galaxies in distant clusters are therefore on
average closer to each other than galaxies in local clusters.

2) The shift to higher individual galaxy number densities at higher
redshifts is consistent with the globally higher number of galaxies
per unit volume on the cluster scale that we obtain
from the way the cluster region is defined and from the
simulation results. The number of galaxies per unit
volume is higher by approximately a factor 1.8 at $z=0.6$, and a
factor 4.7 at $z=1.5$, compared to $z=0$.

3) We find that the global number density $N/V$ evolves because of the
combination of two effects: the DM mass per unit volume ({the mass
density}) increases at higher redshift, simply because the cluster
$R_{200}$ radius is defined to enclose a density that is 200 times the
critical density, while simulations show that the average number of
galaxies per unit of DM mass $N/M$ is constant both with redshift and
cluster mass, being always $\sim 20 \, \rm gal/10^{14} M_{\odot}$
counting only galaxies brighter than a passively evolving $M_V =
-19.4$ at $z=0$.  

The constancy of $N/M$ is found when considering the DM halo mass, but
does not persist when using sim-observed masses derived from the velocity
dispersion as it is done observationally, because the latter strongly
underestimate the system mass at low masses, and slightly overestimate
it at high masses.  In the case of sim-observed masses derived from velocity
dispersions, simulations show a decline of $N/M$ with mass, in
agreement with the observed relations both at high- and low-z.

4) 
From the previous point, it stems that both the mass density and the
global number density evolve as the critical density and do not depend
on cluster mass, therefore they are the same for clusters of all
masses, at a given redshift.  Moreover, the distribution of physical
number densities obtained from the simulations does not vary strongly
with cluster mass.  Hence, contrary to the most intuitive belief of
more massive systems being denser, mentioned in many observational
studies, the most massive and the least massive clusters or groups are
on average equally dense, at a given redshift.

It is important to keep in mind that these conclusions
apply to {\it virialized regions} and for the density estimators
adopted in this paper, that is for a method using the 10 (and, we have
verified, also 3) closest neighbours. These conclusions cannot be
blindly applied to other types of density estimates that probe
physically different scales, nor can be expected to hold in
unvirialized regions of the Universe. For example, the observed shift
in 3D densities with redshift cannot simply be assumed to be valid
also when density is measured as number counts within a 8 Mpc sphere
or a 2Mpc$\times$2Mpc$\times$500$\rm km \, s^{-1}$ cylinder in a
general field survey.

Moreover, we note that in principle these results may depend on the 
semi-analytic model employed, but this can only be ascertained
repeating the analysis with other models.

Our results highlights two main aspects:

a) the strong evolution in the average density conditions experienced by
galaxies in clusters. More distant systems are much denser
environments, both in number and in DM ($\sim$ total)mass.
 
b) the impressive homogeneity in the density of clusters regardless of
mass, at a given redshift. Not only the DM mass density, but also the
average number density of galaxies and, approximately, the distribution
of physical number densities are the same in the virialized regions of
clusters and groups of any mass, at a given
redshift. Cluster-to-cluster variations {\it at a given mass},
however, may be important for individual systems, especially at masses
below $10^{14} M_{\odot}$.

\section*{Acknowledgments} 
BMP thanks the Alexander von Humboldt Foundation and
the Max Planck Instituut fur Extraterrestrische Physik in Garching for
a very pleasant and productive staying during which the work presented
in this paper was carried out.  The Millennium Simulation data bases
used in this paper and the web application providing online access to
them were constructed as part of the activities of the German
Astrophysical Virtual Observatory.  BMP acknowledges financial support
from ASI contract I/016/07/0. GDL
acknowledges financial support from the European Research Council under the
European Community's Seventh Framework Programme (FP7/2007-2013)/ERC grant
agreement n. 202781.

\clearpage

\clearpage

\clearpage

\clearpage

\clearpage

\clearpage

\clearpage

\clearpage

\clearpage

\end{document}